# VO$_2$ Phase Change Electrodes in Li-ion Batteries

Samuel Castro-Pardo[1,2], Anand B. Puthirath[1]*, Shaoxun Fan[1], Sreehari Saju[1], Guang Yang[3], Jagjit Nanda[3,4], Robert Vajtai[1], Ming Tang[1], Pulickel M. Ajayan[1]*.

[1] Department of Materials Science and NanoEngineering, Rice University, 6100 Main St., Houston, TX 77005, USA

[2] Department of Chemistry, Rice University, 6100 Main St., Houston, TX 77005, USA

[3] Chemical Sciences Division, Oak Ridge National Laboratory, 1 Bethel Valley Rd, Oak Ridge, TN 37830

[4] SLAC National Lab and Stanford University, 2575 Sand Hill Road, Menlo Park, CA 94025, USA

*Corresponding Author: anandputhirath@rice.edu  and ajayan@rice.edu

**Abstract**

Use of electrode materials that show phase change behavior and hence drastic changes in electrochemical activity during operation, have not been explored for Li-ion batteries. Here we demonstrate the vanadium oxide ($VO_2$) cathode that undergoes metal-insulator transition due to first-order structural phase transition at accessible temperature of 68°C for battery operation. Using a suitable electrolyte operable across the phase transition range and compatible with vanadium oxide cathodes, we studied the effect of electrode structure change on lithium insertion followed by the electrochemical characteristics above and below the phase transition temperature. The high-temperature $VO_2$ phase shows significantly improved capacitance, enhanced current rate capabilities, improved electrical conductivity and lithium-ion diffusivity compared to the insulating low temperature phase. This opens up new avenues for electrode designs, allowing manipulation of electrochemical reactions around phase transition temperatures, and in particular enhancing electrochemical properties at elevated temperatures contrary to existing classes of battery chemistries that lead to performance deterioration at elevated temperatures.

**Keywords**: Li-ion battery, Phase Changing Materials, Metal insulator transition, Vanadium dioxide

**Introduction**

In the continuous urge to reduce energy production from fossil fuels, green alternatives such as solar, wind, tidal energy are being considered as the best alternatives but in a limited way due to their intermittent nature and comparatively low power delivery that warrant innovation in energy storage technologies to ensure power management and ensure all time deliverability. Owing to the high energy density, lightweight, low self-discharge rate, long shelf life, and stability, Li-ion batteries (LIBs) are the most important and versatile technology thus far for portable electronics and electric vehicles. Traditionally, LIBs are operated in the ambient temperature range and there is a quest for technologies and materials with a wider temperature operation range (>60 ºC) due to applications such as military-grade batteries, electric vehicles, subsea exploration, thermal reactors, space vehicles, and medical devices.[1]

State-of-the-art primary and rechargeable cells are limited in both capacity (150-250 mAh/g) and temperature range (-25°C to 60°C) and is a problem for a number of applications that require high energy rechargeable batteries that operate at extreme environments. Going above the maximum operating temperature risks degradation and irrecoverable damage, often resulting in reduced cell capacity, reduced cell lifetime, cell failure, and in some cases, fires and explosions. Currently, the options for high-temperature lithium-ion secondary batteries are limited due to the instability of the electrode-electrolyte interfaces, dissolution of electrodes, poor electrical conductivity, and finally, the electrolyte itself. Hence, the development of lithium-ion batteries suitable for high-temperature applications requires a holistic approach to battery design because degradation of some of the battery components can produce a severe deterioration of the other components, and the products of degradation are often more reactive than the starting materials.

Among various other layered oxide electrodes currently sought after available, Vanadium-based oxides are considered as one of the most promising electrode materials for next-generation advanced electrochemical energy storage technology due to their layered-crystal structural, high specific capacity, abundance and low cost.[2] The vanadium oxide family is composed of more than 20 stable and metastable compounds with interesting properties including first order phase transition at accessible temperature ranges, that vary depending on the oxygen content in their crystal structure. Among others, Vanadium dioxide ($VO_2$) is particularly promising for lithium-ion batteries applications due to the high specific gravimetric capacity (323 mAh/g for 1 equivalent of Li), and a wide electrochemical potential window (1.5 to 4 V) compared to traditional cathodes such as Nickel Manganese Cobalt (NMC), Lithium Iron Phosphate (LFP), and Lithium Cobalt Oxide (LCO).[3] $VO_2$ can be synthesized in four polymorphs named $VO_2$(A), metastable (B), (M), and (R). Among other, $VO_2$(B) is the most studied material as an intercalation/insertion cathode due to the layered structure it presents and the feasibility of its synthesis.[4] [5] Manthiram et. al., reported the synthesis and electrochemical evaluation of $VO_2$(B) as an insertion cathode in LIBs. This study showed the reversible electrochemical lithiation of $VO_2$(B) in the range of 1 to 3.5 V vs. Li/Li$^+$ with a specific gravimetric capacity of 325 mAh/g corresponding to the insertion of 1 Li per $VO_2$ unit.[6] However, as a metastable monoclinic phase, $VO_2$(B) can be irreversibly transformed to the monoclinic (M) or rutile (R) phases under heat treatment.[6,7] Corr et. al., studied the transformation of the $VO_2$(B) metastable phase to the rutile phase when heated in a Nitrogen atmosphere and monitoring using *in situ* thermodiffractometry technique. The results suggested that the transformation starts at 450 °C but the complete transformation occurred at a temperature of 550 °C.[8] Further studies have shown that the irreversible transformation can occur in the temperature range of 300-600 °C under various types of atmospheres including $N_2$, Ar, and under vacuum.[9,10] $VO_2$(M) with a monoclinic crystal structure, is of a great interest in applications as an optical switch, sensor[11], smart window[12], and optical memory device due to the modulation of its electrical and optical properties upon the application of an external stimulus, i.e., heat.[13] However, in contrast to $VO_2$(B), very few reports have been published studying the electrochemical properties of the $VO_2$(M) phase as well as the high temperature $VO_2$(R) phase as Li-insertion cathodes.

Here in we investigated the vanadium oxide ($VO_2$) system as a positive electrode material that undergoes a metal-insulator transition (MIT) as a consequence of the first-order structural transition from monoclinic to tetragonal at an accessible temperature (68°C). We started the investigation by modelling the room temperature and high temperature phases of $VO_2$ and calculated the difference in the Li insertion sites and variation in the lattice parameters during the event of Li intercalation from zero fraction to one. After identifying a suitable electrolyte operable at elevated temperatures and compatible with vanadium oxide cathodes, we have studied the effect of MIT in Li insertion characteristics followed by the electrochemical characteristics at room and high temperatures. The high-temperature phase not only showed improved capacitance (>70%) but enhanced current rate capabilities and enabled improved active materials loading owing to the high electrical conductivity, improved Li-ion diffusivity, and stability, and open a new avenue of electrode materials that enhance the electrochemical properties at elevated temperatures contrary to the existing class of battery chemistry that deteriorate the performance at high temperatures.

**Results and discussion**

**Simulation of the $VO_2$ (M) and $VO_2$ (R) Phases for Li ion insertions chemistry**

To start with, DFT calculations were performed for both $VO_2$(M) phase and $VO_2$(R) phase to analyze the band structure, density of states, identify the Li occupying sites on intercalation event and how the lattice parameter varies when the state of charge (SoC) changes during the course of Li intercalation from zero fraction to one. It is well known that $VO_2$(M) material undergoes a first order phase transition forming a metastable rutile crystal system $VO_2$(R) with a tetragonal unit cell in a reversible fashion accompanied by an increase in the electrical conductivity by 3 to 5 orders of magnitude.[14] This reversible transformation, so-called metallic-to-insulating transition (MTI) occurs at around 68°C. $VO_2$(R) has a tetragonal rutile structure with $V^{4+}$ ions occupying the corner and center positions, and each $V^{4+}$ is surrounded by six $O^{2-}$. The closest V–V distance in chains along the c-axis is around 2.85 Å. In contrast, the crystal structure of $VO_2$(M) is a distorted rutile structure of $VO_2$(R), with the unit cell of $VO_2$(M) doubling compared to $VO_2$(R). Two different V–V distances equivalent to 3.19 and 2.60 Å between the nearest vanadium atoms form zigzag atom chains.[15] The electronic properties of these two polymorphs change upon the MIT.

Schematic showing the variation properties that augment the Li insertion characteristics such as electric conductivity, Li ion diffusivity and number of Li occupying sites available for both $VO_2$(M) and $VO_2$(R) and voltage discharge profile is displayed in **Figure 1(a-h) and Figure SI1.** The monoclinic phase electronic structure shows the character of a semiconductor material with a narrow indirect bandgap of 0.45 eV. On the other hand, in the rutile phase, the Fermi level ($E_F$) overlaps a manyfold of bands with a predominant $V_{3d}$ character thus giving the material a metallic behavior (no bandgap is obtained). The calculated density of states is displayed in Error! Reference source not found.**c & e** respectively. In the monoclinic crystal structure, V atom is coordinated by six O atoms forming zigzag (**Figure 1a**) empty channel along c direction where Li atoms can be inserted. While, in the rutile crystal structure, each V atom is coordinated by six O atoms forming empty channels along c direction (Error! Reference source not found.**b**) which can be inserted by Li atoms. In the monoclinic phase there are two site per unit cells for inserted Li atoms: 4c sites (0.5, 0, 0) and 4d sites (0.5, 0, 0.25) (**Figure 1g**), similarly there are two sites for the inserted Li atoms: 4c sites (0.5, 0, 0) and 4d sites (0.5, 0, 0.25) (Error! Reference source not found.**h**). The latter is energetically more favorable, according to the calculations. The change of lattice constants on intercalation of Li at various SoC (0-0.2) is shown in **Figure 1d & f** respectively. It can be seen that the lattice constants along b and c are showing increasing trend for VO$_2$(M) and along a and b directions for VO$_2$(R) phase**.** And along a and c direction, lattice constant decreases respectively for M and R phases respectively. From the calculation results it can be concluded that, VO$_2$(M) phase is offering better ecosystem for Li intercalation process in terms of the availability of more Li occupying sites, improved inter lattice diffusivity owing to the presence of linear empty channels along c direction and improved electrical conductivity that augment the overall kinetics of redox reactions.

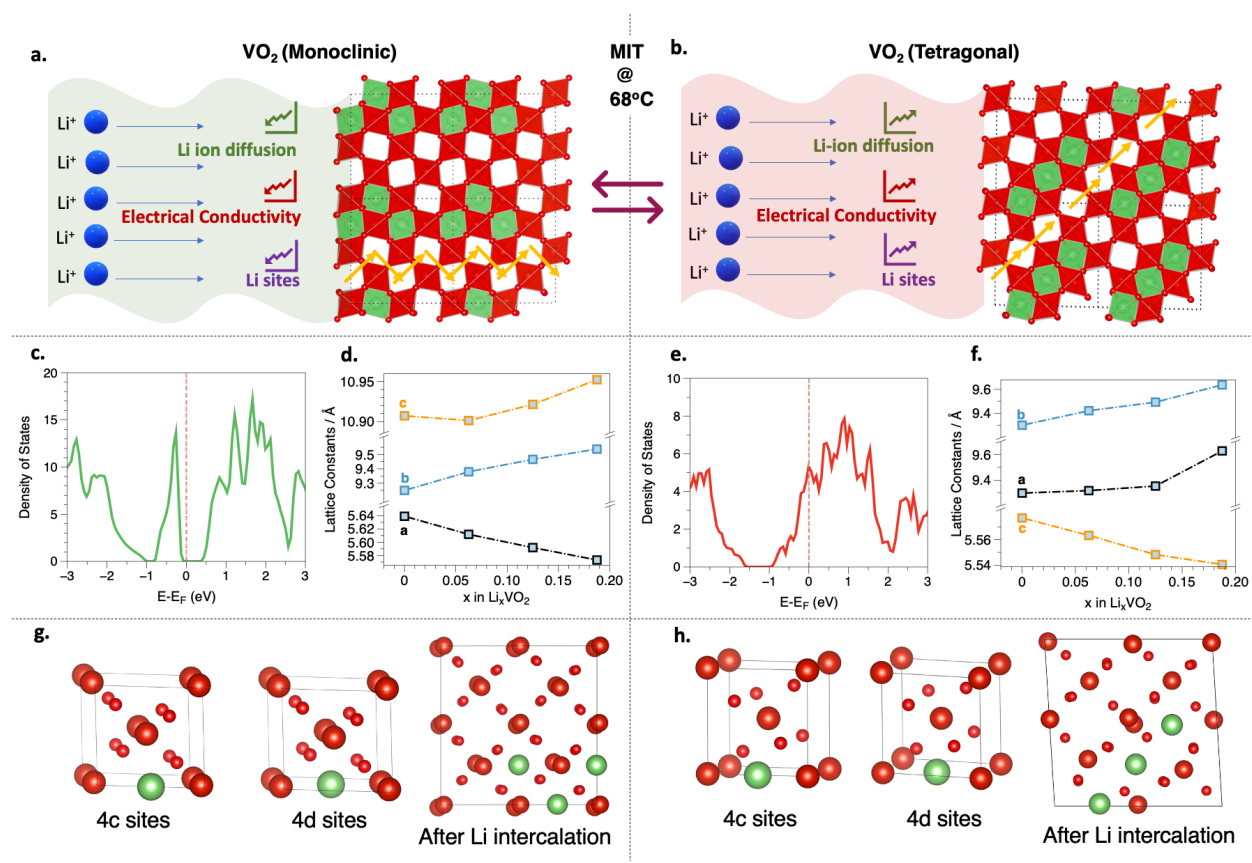

**Figure 1. Simulation of the crystal structures and their variation on the event of Li insertion.** The orientation of empty sites available for Li insertion (a) zigzag (yellow arrows) for VO$_2$(M) phase and (b) linear (yellow arrows) for VO$_2$(R) Phase (c & e) the density of states of VO$_2$(M) and VO$_2$(R) phase and (d & e) the variation of lattice constants with the intercalation of the Li into the corresponding lattice respectively. (g & h) Li occupation sites possible in unit cells and the orientation of intercalated Li ions in VO$_2$(M) and VO$_2$(R) phase lattices respectively.

**The synthesis and structure evolution of VO$_2$(M) Phase from VO$_2$(B) Phases**

Observing the improved crystallographic properties favoring enhanced Li insertion kinetics and chemistry, we next experimented the electrochemical properties of the VO$_2$(M) and VO$_2$(R) phases by reversibly ramping the temperature between 25°C and 75°C that spans across the MIT transition window. The investigation starts with the synthesis of ultrapure VO$_2$(M) phase from its stable oxide V$_2$O$_5$. Phase purity of the VO$_2$(M) is extremely important here as the vanadium in vanadium oxides compounds have tendency to possess mixed valency states that could jeopardize the experiment results and inference thereof. The VO$_2$(M) sample synthesis protocols initially from V$_2$O$_5$ and later from VO$_2$(B) are detailed in the materials and methods section and schematically presented in **Figure 2a.** The crystal structure and chemical composition of the synthesized materials were characterized during the optimization of the synthesis route. The structure evolution of VO$_2$ (M) phase from VO$_2$ (B) phase is detailed here with the help of structural, elemental and differential scanning calorimetry analysis methods. The synthesis of the VO$_2$(B) was optimized by varying the reaction time to achieve the material with the highest crystallinity to ensure a diffusion of Li$^+$ and a high reaction yield. Preliminary XRD patterns were taken of the samples synthesized at 3, 6, 12, 18, and 24 h. To study the progression in the synthesis of VO$_2$, the peak at 20.3° corresponding to the (110) main crystallographic plane in V$_2$O$_5$ was monitored.[16] **Figure 1b** shows the powder XRD traces of the vanadium precursor and synthesized compounds. It was found that the V$_2$O$_5$ precursor started to be consumed after 3 hours of reaction, but its presence was still detected in the reaction products at 3, 6 and 12 hours. At reaction times of 18 and 24 hours, the precursor was not detected. Elemental XPS analysis (**Figure 1c**) was obtained in the range from 510 to 540 eV that corresponds to the core V2p core level signals and O1s.[17,18] Signals located at 516 and 517 eV belong to the V2p$_{3/2}$ species, specifically V$^{4+}$ and V$^{5+}$.[19] Comparing the signal in the range of 516-517 eV it can be noticed that the sample at 3 hrs. of synthesis is composed of a mixture of vanadium oxidation states due to an incomplete reaction. This result is consistent with the XRD pattern in Error! Reference source not found. **1b**. The content of V$^{5+}$ decreased in the samples at 6 and 12 h. of synthesis to less than 5 wt. % and no V$^{5+}$ oxidation state was detected by XPS in the samples at 18 and 24 h. of synthesis. The signals located in the range of 525-524 eV corresponding to V2p$_{1/2}$ state also support the conversion of the V$^{5+}$ to V$^{4+}$ during the synthesis by shifting the maximum signal to a lower binding energy due to an increase in the number of the 3d electrons.[18] In vanadium oxide materials, the presence of two peaks in the 2p region is attributed to the spin-orbital splitting.[20] Quantitative analysis showed that 40 wt. % of the sample is composed of $V^{5+}$ for 3hr processed sample and 0% for 18 and 24 hr samples. Hence, the reaction time of 24 hours was selected to further synthesize the $VO_2$(B) compound since it provided the highest yield of reaction (0.6 g, 48.2%) with at most $VO_2$(B) phase purity.

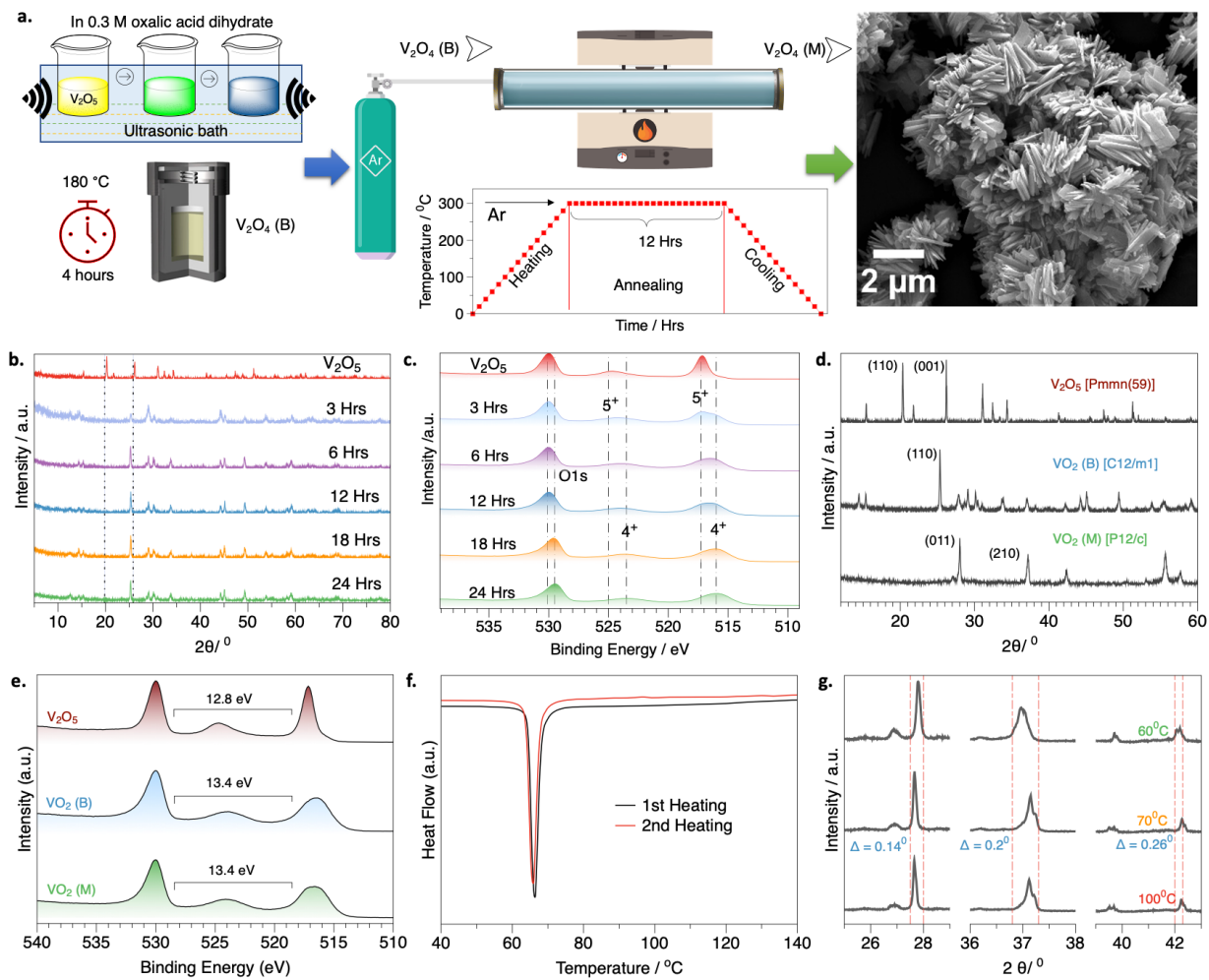

**Figure 2. Synthesis and structural evolution of $VO_2$(M) and $VO_2$(R) phases** (a). Synthesis protocol and SEM images of $VO_2$(M) powder sample (b & c) XRD patterns and High-resolution XPS analysis of the O1s and V2p signals of vanadium precursor and synthesized compounds. (d & e). The XRD spectra of $V_2O_5$, $VO_2$(B) and $VO_2$(M) and XPS and high-resolution spectra of V2p signals. (f). DSC curves showing the reversible monoclinic to tetragonal transition at 66°C. (g). XRD patterns showing the monoclinic to tetragonal transition.

The powder X-ray diffraction (XRD) spectra of the vanadium precursor and optimized synthesized compounds are displayed in **Figure c**. The $V_2O_5$ XRD pattern shows its main diffraction peaks at 26.2° and 20.3° attributed to the crystallographic planes (110) and (001), corresponding to the alpha phase of $V_2O_5$. It belongs to an orthorhombic crystal system with a Pmmn 59 space group, its crystallographic arrangement can be described as a-b layers composed of distorted trigonal bipyramidal coordination polyhedra of O atoms around V atoms.[17]. The strong intralayer forces given by covalent bonds in the $V_2O_5$ pyramids and the weak interlayer interactions facilitate the intercalation of ions between the interlayers along the a-b plane. [21,22] The as-synthesized 24-hour vanadium oxide showed the main diffraction peak at 25.3° attributed to the (110) crystallographic plane of $VO_2$(B). The $VO_2$(B) material possesses a monoclinic crystal system with a space group C12/m1.[23] Its crystal structure is composed of 010 domains where the V atoms are arranged in a parallelogram separated by O planes.[24] $VO_2$(B) is formed by distorted $VO_6$ octahedral units that share corners and edges leading to the formation of tunnels in the a, b and c direction that allows and easy yet reversible intercalation of $Li^+$.[25] It has a unit cell with parameters of a: 12.06 Å, b: 3.69 Å, c: 6.42 Å, α=γ: 90.0°, β: 106.9°, and a cell volume of 273.41 Å³ (PDF# 04-014-1695).[16,26]

Later, to prepare the $VO_2$(M), the as-synthesized $VO_2$(B) was annealed at 300 ºC for 12 h under an Ar-flow to avoid re-oxidation. In all the cases, oxygen needed to be eliminated by either vacuuming the system or purging argon gas for long periods of time since $VO_2$ gets oxidized easily to $V_2O_5$. The $VO_2$(M) shows a monoclinic crystal system with a P21/c space group. The XRD pattern of the annealed sample shows the main diffraction peaks at 28.0º and 37.1º due to the crystallographic rearrangement of the crystal system. These peaks correspond to the (011) and (210) planes. The unit cell parameters of a: 5.74 Å, b: 4.51 Å, c: 5.37 Å, α=γ: 90.0°, β: 122.6°, and a cell volume of 117.46 Å³ (PDF# 00-009-0142).[27] Within the crystal lattice, the $V^{4+}$ is bonded to six $O^{2-}$ atoms to form a mixture of edge and corner-sharing $VO_6$ octahedral units. There is a spread of V-O bond distances ranging from 1.80-2.06 Å and V-V distances of 3.2 and 2.60 Å.

XPS elemental analysis was performed on the powder samples to confirm no re-oxidation occurred during the thermal annealing of the $VO_2$(B) material. The difference in the binding energy between the O1s core level, located at 530 eV, and the V2p3/2 level (ΔE) was calculated as shown in **Figure e.** The ΔE for $V_2O_5$ was 12.8 eV whereas 13.4 eV for the $VO_2$ samples, confirming the presence of the $V^{5+}$ and $V^{4+}$ states, respectively. It is well documented that the ΔE increases as the oxidation state of vanadium decreases[28] and the V2p3/2 signal becomes broader as the oxidation state decreases due to an increase in the number of available multiplet configurations in the photoemission final states, relating to the coupling of the core hole to the 3d electron.[29] The particle morphology and size were investigated by SEM. As seen in **Figure a**, the VO$_2$(M) was synthesized in a prismatic shaped particles with average dimensions of 1 μm in length, 300 nm in width, and 100 nm thick. The morphology of VO$_2$(B) particles was maintain upon the irreversible transformation.

DSC experiments were performed to reveal the transformation of VO$_2$(M) to VO$_2$(R) upon heating. Thermograms in **Figure f** shows a first-order phase transition in the VO$_2$(M) material as indicated by an endothermic signal occurring at 66ºC. This endothermic feature is consistent with previous reports;[30] the so-called metallic-to-insulating transition (MTI), is expected to occur in the temperature range of 66-68 ºC. The synthesized VO$_2$(M) shows a reversible MTI transition as seen by comparing the 1$^{st}$ and 2$^{nd}$ heating curves. XRD traces shown in **Figure g** were obtained at different temperatures to show the main crystallographic changes upon the MTI. There is a clear indication of the transformation of the monoclinic phase to the tetragonal phase by comparing the main (011) and (210) planes. At 60ºC, the VO$_2$(M) shows the main (011) and (210) planes at 27.79 and 36.97º, respectively. These values appeared to shift to 27.66 and 37.15 at 70ºC due to the transformation to the tetragonal form. During this reversible transformation, the V-V chains rearrange from a zigzag configuration in the distorted monoclinic phase to a V-V linear arrangement in the rutile phase causing a displacement in the crystallographic planes. A further increase to 100 ºC only causes a small shift due to the thermal expansion of the material. In addition, the signal located at 37º, after the MTI, shows a shoulder that appears at 37.23º and eventually shifts to 37.21º due to thermal expansion. In the same fashion, after the MTI, one new signal appears at 39.52º and four more disappear, signals located at 42.07º, 33.37º, 44.06º, and 48.38º, confirming the transition upon the application of a heat stimulus.

**Electrochemical Performance of the Li-ion cells at VO$_2$(M) and VO$_2$(R) Phases**

Later, we have studied the electrochemical performance of the cells assembled with VO$_2$ (M) phase as positive electrode by ramping the cells across room temperature and 75 degrees as it allows to study the cell performance at both RT phase as well as high temperature phase. High temperature grade coin cell components were used to ensure the stability of the cells up to 150ºC. In the study of the electrochemical evaluation of VO$_2$ phases, preliminary experiments using the traditional LiTFSI in EC: DMC electrolyte suggested a poor electrochemical activity with a fast capacity fade in the rutile VO$_2$ (**Figure S2**). Post-mortem XPS and Raman analysis showed that VO$_2$(M) was partially dissolved in the electrolyte, and the increase in the temperature to produce the rutile phase further accelerated the dissolution process (**Figure S3**). The electrochemical properties of the optimized VO$_2$(M) and VO$_2$(R) were investigated in a series of experiments, including by employing a solid polymer electrolyte (**Figure S4**) that proved the stable reversible electrochemical lithiation of the crystals on using the 1M LiTFSI in [EMIM: TFSI]: PC ionic liquid electrolyte.

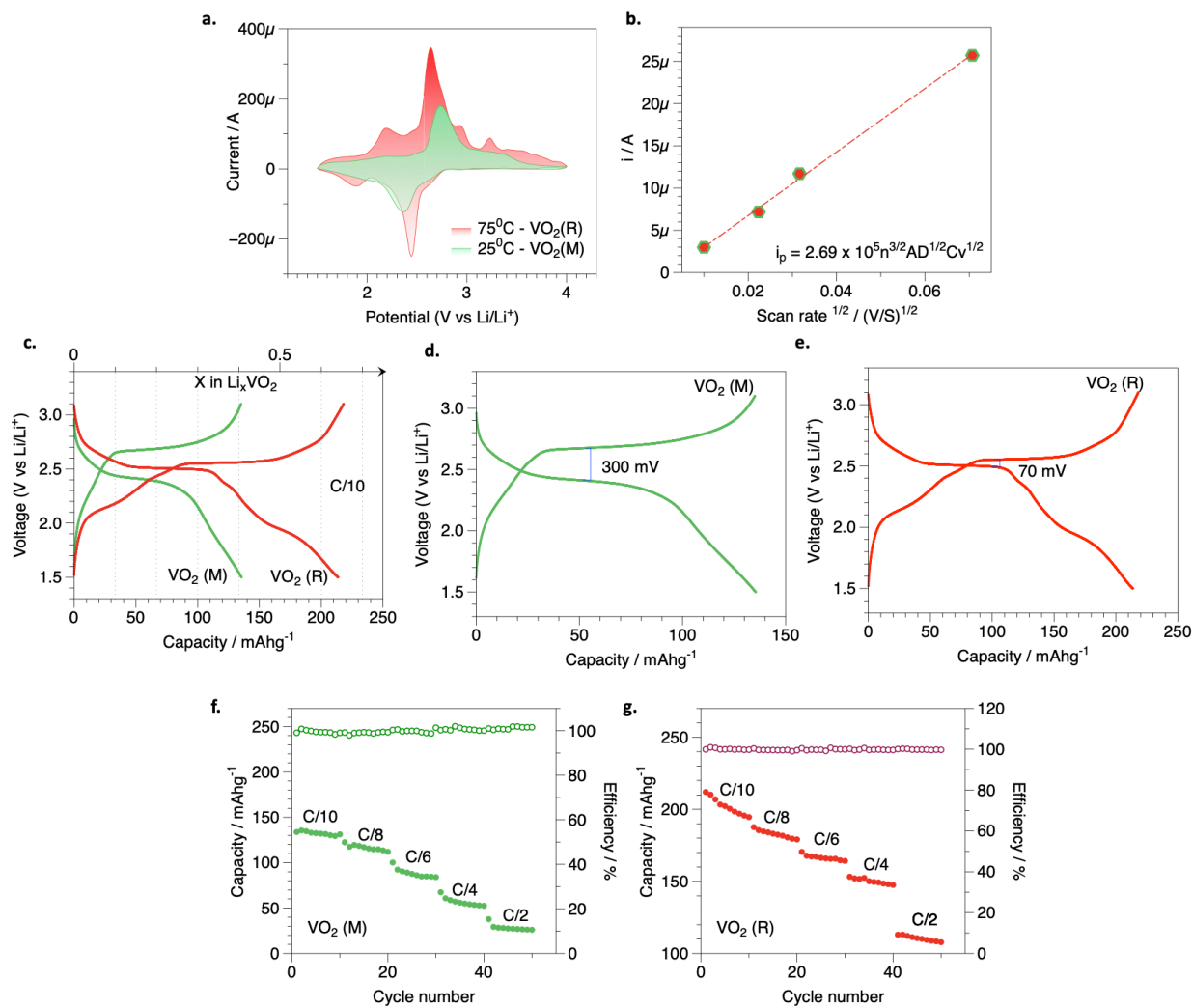

**Figure 3. Electrochemical evaluation of VO$_2$(M) and (R) phases.** (a). Cyclic voltammetry of a VO$_2$/Li cell at a scan rate of 0.1 mV/s. (b) Randles-Sevcik relationship for the de-intercalation processes in VO$_2$(M). (c-e). Galvanostatic charge-discharge curves of VO$_2$(M) and VO$_2$(R) and filling fraction of Li in VO$_2$. (f-g). Long cycling performance at a current rate of C/10 and C/8 (top) and rate performance at increasing C rates (bottom) respectively for VO$_2$(M) and VO$_2$(R)

Cyclic voltammetry experiments were performed using both systems at a scan rate of 0.1 mV/s. The cyclic voltammogram (**Figure a**) corresponding to the monoclinic phase shows one pair of redox peaks located at 2.36 and 2.73 V vs. Li/Li$^+$ in the cathodic and anodic region, respectively. The cathodic peak is attributed to the lithiation of $VO_2(M)$ to $Li_xVO_2$ with a lithium content of x<0.5 based on the calculation of the charge under the peak. The anodic peak is attributed to the delithiation step in the reverse reaction. The peak separation (anodic-cathodic overpotential or hysteresis) of about 370 mV indicates a high internal resistance during the intercalation/deintercalation process.[31] By varying the scan rate of the CV experiment and obtaining a linear relationship between the peak current vs. the square root of the scan rate (**Figure 3b**), it can be concluded that the insertion (lithiation/delithiation) processes happening in the $VO_2(M)$ material follow an electrochemically reversible electron transfer process involving freely diffusion redox species governed by the Randles-Sevcik diffusion-controlled mechanism.[32] An estimated diffusion coefficient of $Li^+$ was obtained from the Randles–Sevcik equation and tabulated in **Table S1.** The diffusion coefficient values for the cathodic and anodic peak current in the voltammogram shown in **Figure a** are $1.4 \times 10^{-13}$ and $3.0 \times 10^{-13}$ $m^2/s$. In the case of the $VO_2(R)$, the cyclic voltammogram in **Figure a** shows two redox pairs, one located at 2.44/2.63 and the second one in the potentials of 1.88/2.18 V vs. $Li/Li^+$. Two reduction processes observed in the cathodic direction are located at 2.44 and 1.88 V vs. $Li/Li^+$. The cathodic peak located at 2.44 V vs. $Li/Li^+$ is consistent with the cathodic peak in $VO_2(M)$ which is attributed to the insertion of $xLi^+$ in the crystal lattice forming a solid solution of $Li_xVO_2$. The second cathodic peak located at 1.88 V vs. $Li/Li^+$ corresponds to the subsequent lithiation of $Li_xVO_2$ to a solid solution with higher content of $Li^+$. In the reverse direction, the delithiation goes from a solid solution of higher Li content to a solid solution of lower Li content at a potential of 2.18 V vs. $Li/Li^+$ and a further fully delithiation step at a potential of 2.63 V vs. $Li/Li^+$. Two lithiation and delithiation potentials indicate two different active sites in the solid solution of $VO_2(R)$ or the rearrangement of Li ions within the crystal lattice (hopping mechanism). It can be concluded that the rutile phase has more active sites for Li insertion than the monoclinic phase. Both redox processes follow the reversible electrochemical diffusion relationship of Randles-Sevcik. In both cases, cathodic and anodic processes, the diffusion coefficients of $Li^+$ are higher in the case of $VO_2(R)$ than in $VO_2(M)$. This is attributed to higher affinity towards the rutile phase, an easier intercalation/deintercalation that could be due to an increment in the temperature of the system to achieve the rutile phase and a higher conductivity of the rutile phase. The smaller hysteresis values in $VO_2(R)$(180 mV) compared to $VO_2(M)$(370 mV) indicate a lower intercalation/deintercalation resistance during the electrochemical processes. An extra peak at 3.22 V vs. $Li/Li^+$ located in the anodic direction indicates the transformation of the $VO_2$ to $V_2O_5$ in an irreversible fashion since there is no peak in the cathodic direction. Therefore, it is concluded that the electrochemical window for a reversible lithiation/delithiation in $VO_2(R)$ should be limited between 1.5 to 3.1 V vs. $Li/Li^+$. The $VO_2(R)$ material shows higher electrochemical activity than the $VO_2(M)$ since the rutile phases shows higher current density and the area under the curve than the monoclinic phase. Both materials follow a reversible electrochemical process during lithiation and delithiation but there is a clear indication of a higher electrochemical activity in the rutile phase compared to the monoclinic phase.

The cycling performance of the two materials was hence studied in the electrochemical window potential of 1.5 to 3.1 V vs. $Li/Li^+$ to avoid oxidation of $VO_2(R)$ to $V_2O_5$. Galvanostatic experiments were performed to identify the specific gravimetric capacity of the materials and their performance upon cycling at different current densities (C rates). The galvanostatic charge-discharge curve of $VO_2(M)$ at a current density equivalent to C/10 (32.3 µA/mg) in the electrochemical window of 1.5 to 3.1 V vs. $Li/Li^+$ shows an initial specific gravimetric capacity of 135.5 mAh/g according to the **Figure c**. The discharge curve is characterized by a sloping plateau with a mid-voltage of 2.43 V vs. $Li/Li^+$, which indicates the intercalation of $Li^+$ to the crystal structure of $VO_2(M)$. The highest contribution in capacity occurs in this sloped plateau due to the transformation of $V^{4+}$ species to $V^{3+}$ species. The reverse process during the charge step follows the same sloped plateau but with a higher mid-voltage of 2.73 V vs. $Li/Li^+$, indicating an overpotential or hysteresis value of 300 mV, **Figure d**. This hysteresis value is consistent with the value obtained by cyclic voltammetry in **Figure a**. The filling fraction (**Figure c**) of the $VO_2(M)$ system was calculated in the same electrochemical conditions using a current density equivalent to C/10. The $VO_2(M)$ system shows a filling fraction equivalent to 0.42 Li per $VO_2$ unit, consistent with the prediction based on the cyclic voltammogram (charge under the cathodic peak). To test the rate capability of this system, 10 consecutive charge-discharge cycles were evaluated at different C rates equivalent to C/10 (32.3 µA/mg), C/8 (40.4 µA/mg), C/6 (53.8 µA/mg), C/4 (80.7 µA/mg), and C/2 (161.2 µA/mg). The discharge gravimetric capacities were estimated to be 130, 115, 86, 55, and 27 mAh/g in the ascending order of C rate, as shown in **Figure f**. This system shows good rate capability in slow current rates (C/8 to C/6), but at faster current rates, the capacity faded. The Coulombic efficiency (reversibility during each cycle) remained constant in the 99.1 to 100 % range during the cycling at different current rates. On the other hand, the capacity retention decreased with an increase in the current density. The capacity retention, calculated as the discharge capacity at cycle $10^{th}$ over the $1^{st}$ discharge capacity, was estimated to be 98% at C/10, 95.3% at C/8, 91% at C/6, 86% at C/4, and 89% at C/2. The galvanostatic charge-discharge curve of $VO_2(R)$ at a current density equivalent to C/10 (32.3 µA/mg) in the electrochemical window of 1.5 to 3.1 V vs. $Li/Li^+$ shows an initial specific gravimetric capacity of 210 mAh/g according to the **Figure c**. The discharge curve is characterized by a sloping plateau with a mid-voltage of 2.49 V vs. $Li/Li^+$, which indicates the intercalation of $Li^+$ to the crystal structure of $VO_2(R)$. The highest contribution in capacity occurs in this sloped plateau due to the transformation of $V^{4+}$ species to $V^{3+}$ species during the formation of the $Li_XVO_2$. A second sloped plateau is observed at a mid-voltage of 1.9 V vs. $Li/Li^+$, which is consistent with the value of 1.88 observed in the cyclic voltammogram upon an extra lithiation step. The reverse process during the charge step follows the same sloped plateaus with mid-voltage of 2.2 V and 2.50 V vs. $Li/Li^+$, indicating a hysteresis value of 70 mV in the later plateau, **Figure 3e.** This hysteresis value is lower than the value obtained by cyclic voltammetry. Still, this discrepancy is attributed to the different experimental conditions used in these experiments (constant voltage vs. constant current). The decrease in hysteresis in the $VO_2(R)$ indicates an easier diffusion of Li within the crystal lattice which is also consistent with the results obtained in cyclic voltammetry experiments. The higher electrochemical activity observed in the cyclic voltammetry experiments in terms of higher current density is consistent with the information obtained in the galvanostatic experiments. A higher specific capacity in the discharge curve in the rutile phase compared to the monoclinic phase is translated as a higher filling fraction of Li in $LixVO_2$ with a value of 0.67 Li per $VO_2$ unit, as observed in **Figure c**. The additional features observed in the charge-discharge curves in the rutile phase are also consistent with the potentials of the features/signals found in the cyclic voltammetry experiment, which are attributed to further lithiation/delithiation and ion hoping within the crystal lattice. The discharge gravimetric capacities obtained during rate capability experiments were determined to be 210, 187, 170, 153, and 113 mAh/g in the ascending order of the C rate, as shown in **Figure 3g**. This system shows good rate capability even at faster current rates (C/2) compared to the monoclinic phase. The Coulombic efficiency remained constant in the 99 to 100 % range during the cycling at different current rates. The capacity retention, contrary to the monoclinic phase findings, remained almost constant with an increase in the current density. The capacity retention, calculated as the discharge capacity at cycle $10^{th}$ over the 1st discharge capacity, was estimated to be 92% at C/10, 95% at C/8, 96.3% at C/6, 96.2% at C/4, and 95.4% at C/2. A capacity retention of 96% compared to the 1st discharge cycle could mean a cycling performance up to 60 cycles before the capacity fades to a capacity retention lower than 80%.

**Electrochemical Impedance Spectroscopy and Li diffusion coefficient Determination**

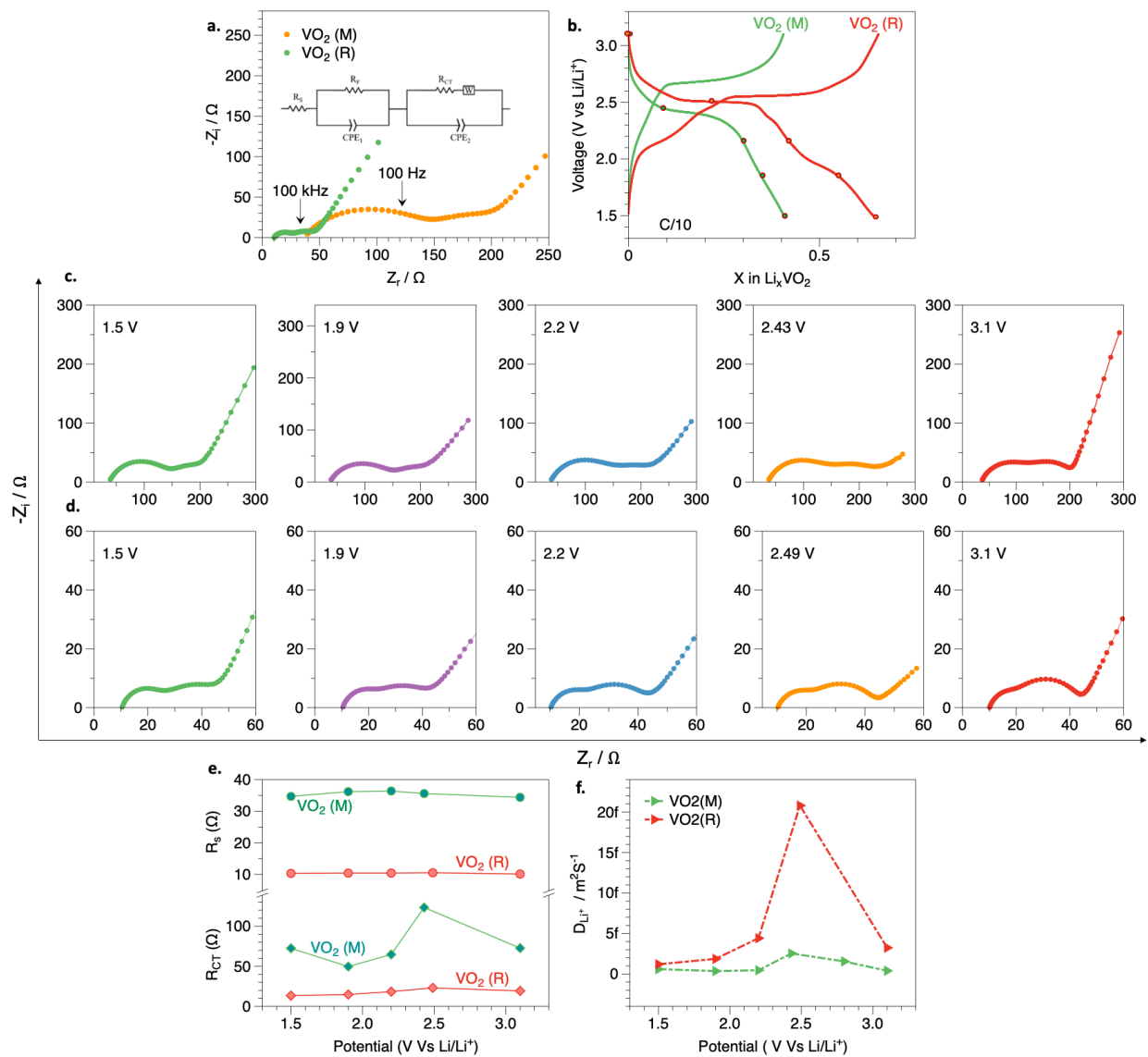

**Figure 4. Electrochemical Impedance Spectroscopy analysis and Li ion diffusion coefficient.** (a)Nyquist plot of $VO_2(M)$ and $VO_2(R)$ based cells at zero bias conditions. The equivalent circuit is shown in the inset. (b) The SoC (shown by dots in the discharge curve) at which the impedance measurements were performed to evaluate the variation of $R_S$ and $R_{CT}$. (c) Nyquist plot at different SoC for $VO_2(M)$ and (d) $VO_2(R)$. (e) The curves showing the variation of $R_S$ (top) and $R_{CT}$ (bottom) for $VO_2(M)$ and $VO_2(R)$ with respect to the SoC. (f) The Li diffusion co-efficient across the lattice of $VO_2(M)$(green) and $VO_2(R)$ (red) respectively.

Electrochemical Impedance Spectroscopy (EIS) experiments were performed to understand the enhancement in terms of an improvement in the electrical properties of the material. Comparing the series resistance and charge transfer resistance (Rs and $R_{CT}$) values for both systems after the 50$^{th}$ charge-discharge cycle, **Figure 4a**, which is an indicator of the electrical and ionic conductivity of the system it can be seen that the value of the bulk resistance of the cell using $VO_2$(M) as a cathode is 3.6 times higher than that of the cell using $VO_2$(R) with Rs values of 36.3 and 10.1 Ω corroborating the idea of an increase in electrical conductivity due to the change to a metallic-like material. Upon 50 charge-discharge cycles, the Rs value of the $VO_2$(M) system increased from 28.1 Ω to 36.3 Ω due to the formation of solid electrolyte interfaces. The same behavior was observed in the $VO_2$(R) system. The Rs value increased from 5.3 to 10.1 Ω from the 1$^{st}$ cycle to the 50$^{th}$ cycle. In the $VO_2$(M), the $R_{CT}$ value was calculated to be 123.4 Ω, while for $VO_2$(R), the value is smaller, about 23.1 Ω. This means a higher electrochemical activity and faster kinetics due to the phase transition and the modulation of the electrical properties.[33] The kinetic diffusion of Li$^+$ was studied by fitting the linear part at lower frequencies to a Warburg[34] element and calculating the chemical diffusion coefficient according to the following equation:

$$D_{Li^+} = (RT/n^2F^2ACW)^2$$

The calculated chemical diffusion coefficient was 2.08x10$^{-14}$ m$^2$/s for the $VO_2$(R), which is one order of magnitude higher than the diffusion value calculated for $VO_2$(M) of 2.54x10$^{-15}$ m$^2$/s. The chemical diffusion coefficient was expected to be higher in the rutile phase since this parameter is directly dependent on the temperature of the system and inversely dependent on the Warburg element. To understand the kinetic diffusion of Li$^+$ within the crystal lattice during its operation, EIS experiments were performed at different potentials during one discharge cycle, as shown in **Figure 4c, d & e**. As expected, the Rs value didn't change at different extents of delithiation[35] under a discharge step at C/10, obtaining fairly constant values between 34.7 and 36.3 Ω for $VO_2$(M) and Rs values in the range of 10.1 to 10.6 Ω for $VO_2$(R)**.** On the other hand, the charge transfer resistance changed upon delithiation. The highest values of $R_{CT}$ occur at the potential of the plateau seen in the charge-discharge experiments, which correspond to the resistance getting higher due to the removal of Li$^+$ from the crystal lattice. It involves a physical extraction and ion hopping other than ion shuttle from electrode-electrode. Higher $R_{CT}$ values were found in the monoclinic phase due to slower kinetic electron transfer processes in the semiconducting material. The chemical diffusion coefficient during delithiation, **Figure 4f**, showed a maximum coefficient value at the same potential of the removal of Li from Li$_x$VO$_2$, which is also consistent with the charge-transfer resistance values obtained in **Figure 4 c & d.** In the lithiation process, the highest diffusion coefficient values were also found to be in the plateau region. This is attributed to the affinity of Li towards the crystal lattice at this specific potential.

**Conclusion**

We have studied the phase changing material VO$_2$ as cathode active materials for LIBs with improved extreme environment operability. The results presented here proved the synthetic route followed to produce vanadium dioxide in the monoclinic (B) phase that was consistent and reproducible. The thermal treatment of VO$_2$(B) resulted in an irreversible transformation to VO$_2$(M) with a monoclinic phase. This (M) phase showed a reversible thermal transition at 66 ºC measured by DSC and further confirmed by XRD using a heating stage due to the transition to a rutile system with a tetragonal crystal structure (R). The electrochemical evaluation of both phases proved the hypothesis that the rutile phase of VO$_2$ is a better performing cathode material than the monoclinic phase. In the cyclic voltammetry experiments, an extra redox pair was observed at 1.88/2.18 V vs. Li/Li$^+$ in the rutile phase suggesting a higher reversible lithiation/delithiation capacity and a decrease in the hysteresis measured as the peak distance in the main redox process and a higher current density showed a superior electrochemical activity in the rutile phase compared to the monoclinic phase. The specific gravimetric capacity of the rutile phase, increased by ~70% because of a higher filling fraction (0.42 to 0.67) attributed to a higher number of active sites for Li intercalation. At the same time, the capacity retention improved and remained constant (95%) at different current densities, leading to a better rate capability. Furthermore, electrochemical impedance spectroscopy experiments revealed the improved electronic properties of the systems and results proved a reduction in the bulk resistance, charge-transfer resistance, and an increment of one order of magnitude in the chemical diffusion coefficient of Li$^+$, suggesting an electronic component to the increase in the specific capacity of the rutile phase compared to its monoclinic counterpart. The transition of vanadium dioxide from the monoclinic phase to the rutile phase proved a higher thermal stability during operation and a higher electrochemical activity. Hence, the results shown here open the possibility of studying phase-changing materials as cathodic (and possible anodic) active materials in batteries with high-temperature operability and extreme environment applications.

**Experimental section**

**Synthesis of vanadium dioxide (M)**

VO$_2$(B) was synthesized by a reduction reaction of vanadium pentoxide following a hydrothermal crystal growth. In brief, 15 mmol of V$_2$O$_5$ (Sigma-Aldrich) were added to a 150 mL of an aqueous 0.3 M solution of oxalic acid dihydrate (Sigma-Aldrich) and kept for reaction assisted in a bath sonicator. The color of the solution changed from yellow to green to dark blue upon the reduction of the vanadium ions from the V$^{5+}$ state to the V$^{4+}$ state in the presence of oxalic acid. After 3 hours of sonication, the reaction liquid was transferred to a 200 mL autoclave (filling fraction = 0.75) and kept for hydrothermal crystal growth at 180 °C for 24 hours. The synthesized powder was filtered and washed with DI water and ethanol and dried in a vacuum oven at 80 °C for 12 hours. VO$_2$(M) was synthesized by thermal transformation of as-synthesized VO$_2$(B) in an Argon atmosphere at 300 °C for 12 hours.

**Preparation of electrodes**

Electrodes were prepared by coating an aluminum foil current collector with a slurry of 70 wt. % VO$_2$, 20 wt. % conductive carbon (Super P, Alfa Aesar), and 10 wt. % polyvinylidene fluoride binder (PVDF, Sigma-Aldrich) dispersed in N-Methyl-2-pyrrolidone (NMP, VWR Chemicals BDH®). Electrodes were dried in a vacuum oven at 80 °C, further punched out in 1.22 cm$^2$ circular electrodes, and stored at room temperature in a vacuum desiccator until further use. The thickness and mass loading of the electrodes was 100 μm and 3 mg/cm$^2$, respectively.

**Electrochemical characterization**

2032 type coin-cells (MTI Corporation) were assembled in an Argon-filled glovebox (MBraun, O$_2$ < 0.1 ppm, H$_2$O < 0.1 ppm) using a VO$_2$ working electrode, a 1M LiTFSI (Sigma-Aldrich) in 1-Ethyl-3-methylimidazolium bis(trifluoromethylsulfonyl)imide (io-li-tec) - Propylene carbonate (Sigma-Aldrich) (IL:PC, 8:2 v/v) ionic liquid electrolyte, and a 0.75 mm thick lithium chip (Alfa Aesar) as counter/reference electrode. The specific capacities and cycling performance were determined by galvanostatic charge/discharge experiments at different C rates (1C=323 mAh/g) in the potential range of 1.5-4.0 V vs. Li/Li$^+$ using a battery cycler (Landt). Cyclic voltammograms were obtained by scanning the potential from the OCP to 1.5 V vs. Li/Li$^+$ followed by a potential swept to 4 V vs. Li/Li$^+$ using an autolab potentiostat (Metrohm). EIS measurements were acquired at the OCP in the frequency range of 100 kHz to 1 Hz using an amplitude of 10 mV and 10 points per decade. The monoclinic and tetragonal phases were characterized at 25 and 70 °C, respectively.

**Characterization methods**

X-ray powder diffraction experiments in the 2θ range of 5–80° were performed in a Rigaku Smartlab diffractometer equipped with Cu-Kα radiation (λ = 1.54049 Å) at a scan speed of 1 degree/min and a step of 0.002 degrees. XRD experiments as a function of temperature were acquired using a heating stage by heating the material at a rate of 10 ºC/min and kept for 1 hour before each data acquisition. X-ray photoelectron spectroscopy was performed on a Phi Quantera XPS spectrometer operated at 50 W and 15 kV using a 200 μm spot size analysis. Survey analysis and high-resolution (elemental) analysis were obtained using a pass energy of 140 and 26 eV, respectively. The data was calibrated based on the O1s peak at 530 eV. Confocal Raman microscopy was studied using an InVia Renishaw Raman spectrophotometer coupled to a Leica DMLM microscope. A 50x objective lens and a 532 nm laser at 0.1% power intensity were used for the spectra acquisition. SEM images were obtained in an FEI Quanta 400 ESEM FEG operated in a high-vacuum atmosphere at 20 kV. The samples were coated with gold (5 nm) using a Denton sputtering system. Differential Scanning calorimetry traces were obtained in a TA Instruments DSC Q20 equipment in the temperature range of 0-100 ºC at a scan speed of 10 ºC/min; samples were sealed in an Ar-atmosphere.

**Acknowledgments**

S. C-P would like to acknowledge to the Consejo Nacional de Ciencia y Tecnología (CONACyT) and the Instituto de Innovación y Transferencia de Tecnología de Nuevo León for the support provided under the Ph.D. scholarship program (CVU 747944); to Dr. Rafael Verduzco and Dongjoo Lee for their support during the DSC experiments; to Dr. Bo Chen and Dr. Jianhua Li for the training provided during the XPS and XRD experiments, respectively.

**Competing interests**

Rice University have filed an invention disclosure (Tech ID No. 2023-015) about the electrochemical behavior of $VO_2$ during phase transformations.

**Data and materials availability**

All data are available in the main text or the supplementary materials.